# A simplification of the vorticity equation and an extension of the vorticity persistence theorem to three dimensions

T. S. Morton

Department of Mechanical, Aerospace & Biomedical Engineering, University of Tennessee Space Institute
411 B.H. Goethert Parkway, Tullahoma, TN 37388, USA

It has been known for more than a century that in two-dimensional (planar) Euler flow, vorticity is conserved along streamlines. In three-dimensions, however, no such result has been established, and this is primarily due to the vortex stretching term in the equation of motion. The vorticity persistence theorem is herein extended to three dimensions. It states that all components of the vorticity tensor normal to the streamline direction are conserved along streamlines in Euler flows, provided the streamline coordinate system has one transverse coordinate on which the velocity does not depend. This extension is accomplished with the aid of a mathematical simplification of the vorticity equation derived for arbitrary coordinate systems. What remains of the nonlinear convective terms in the vorticity equation, after the mathematical simplification, is the Lie derivative of the vorticity tensor with respect to fluid velocity. A coordinate-independent temporal derivative is defined which, when set to zero, expresses either the continuity or vorticity equation (excluding the viscous term), depending upon the argument supplied to it.

## 1. Introduction

It is has been known for more than a century that in 2-D (planar) Euler flow, the vorticity, $\omega$, is conserved along streamlines (Helmholtz 1858). Nearly as well known, however, is the fact that such a result has not been established for 3D Euler flows (see, e.g., Majda & Bertozzi 2002; Shariff & Leonard 1992; Friedlander 2002). In fact, research in the last few decades has shown that in the case of axisymmetric Euler rings, it is the quantity $\omega/R$ (where $R$ is the distance from the symmetry axis), not $\omega$ itself, which persists (Fraenkel 1970, Norbury 1972, Hunt and Eames 2002). It will be shown in §7 that these two Euler flows – 2D planar and axisymmetric – may be unified by stating that all off-trajectory components of the vorticity *tensor* persist along streamlines, not only in Euler rings and not only in axisymmetric flows, but in any three-dimensional flow for which $\nu\nabla^2\boldsymbol{\omega} = 0$.

The restriction of the vorticity persistence theorem to planar flow obviously limits the assumptions that can be made when attempting to solve for general three-dimensional flows analytically. Some numerical solution schemes are limited to two dimensions due, in part, to the computational load and large storage requirements associated with the connection coefficients (Christoffel symbols) inherent in general coordinate systems (Lee & Soni 1997; Wesseling *et al.* 1998). The purpose of this paper is to present a mathematical simplification of the vorticity equation valid in arbitrary coordinate systems and to extend the vorticity persistence theorem to three dimensions.

The vector form of the vorticity equation,

$$\frac{\partial \boldsymbol{\omega}}{\partial t} + \boldsymbol{v}\cdot\nabla\boldsymbol{\omega} - \boldsymbol{\omega}\cdot\nabla\boldsymbol{v} = \nu\nabla^2\,\boldsymbol{\omega}\,, \tag{1}$$

where $\boldsymbol{\omega}$, $\boldsymbol{v}$, and $\nu$ are, respectively, the vorticity, velocity, and kinematic viscosity, employs the dot notation of Gibbs and the boldface font introduced by Heaviside. Implied by this "vector notation" is the claim that if the equation is valid in any inertial coordinate system, it is valid in all such coordinate systems. However, when written in the coordinate form

$$\frac{\partial \omega_i}{\partial t} + v_k \frac{\partial \omega_i}{\partial x_k} - \omega^j \frac{\partial v_i}{\partial x_j} = \nu \frac{\partial^2 \omega_i}{\partial x_j^{\,2}}, \tag{2}$$

the equation is valid only in rectangular coordinate systems. Simply stating that (1) must be independent of coordinate system does not offer a unique meaning to (1) analytically, even if it does conceptually. However, if $\boldsymbol{\omega}$ and $\boldsymbol{v}$ are defined as tensors rather than merely as vectors, (1) may be generalized for any inertial coordinate system (see (3) below). (In fact, it is shown in §6 that (1) may be generalized to non-inertial coordinate systems as well.)

If it is claimed only that $\boldsymbol{\omega}$ is a vector, the constraints it must satisfy are milder than those imposed by assuming a tensor character. For example, the requirements imposed by stating that $\boldsymbol{\omega}$ is an element of a vector space are that the sum of any two elements of the space is again a member of the space, and so forth. It does not follow from these requirements that if (1) is valid in one inertial coordinate system, it will be valid in all such systems. In fact, the various components of a vector need not even have remotely related units. Therefore, while all first-order tensors are vectors, not all vectors are tensors. Even the physical velocity of a fluid particle, which is a vector, is not a first-order tensor. This is precisely because the *physical* velocity does not transform tensorially from one coordinate system to another. The same can be said of the physical vorticity. In order to make the distinction between a first-order tensor and a vector whose components may or may not transform tensorially, the vector

$$\left[\frac{\mathrm{d}R}{\mathrm{d}t},\ R\frac{\mathrm{d}\theta}{\mathrm{d}t},\ \frac{\mathrm{d}z}{\mathrm{d}t}\right],$$

for example, will be understood to be a velocity *vector* in a cylindrical coordinate system (the physical velocity), while

$$\left[\frac{\mathrm{d}R}{\mathrm{d}t},\ \frac{\mathrm{d}\theta}{\mathrm{d}t},\ \frac{\mathrm{d}z}{\mathrm{d}t}\right]$$

is the velocity *tensor* in a cylindrical coordinate system. Both are vectors, but only the latter is a first-order tensor. The same distinction must be made between the physical vorticity vector and the vorticity tensor. The term "contravariant vorticity vector" could be used to refer to the vorticity tensor, but this phrase is slightly cumbersome and forces one to specify variance.

It is shown in §6 that in three-dimensional Euler flow, as in the two-dimensional case, any component of the vorticity tensor not directed along fluid particle streamlines must remain constant along streamlines. The key to the derivation is a mathematical simplification of the nonlinear convective terms in the vorticity equation, which reveals that vortex stretching is closely related to the Christoffel symbols of the streamline coordinate system.

## 2. Simplification of the Vorticity Equation

The vorticity equation can be written in contravariant form for an arbitrary inertial coordinate system as:

$$\frac{\partial \omega^i}{\partial t} + v^k \omega^i_{,k} - \omega^j v^i_{,j} = \nu (g^{pk} \omega^i_{,k})_{,p}, \tag{3}$$

where $g^{pk}$ are components of the inverse of the metric tensor $g_{pk}$ of the arbitrary coordinate system. In (3), $\omega^i$ and $v^i$ are, respectively, the $i$th contravariant components of the vorticity and velocity *tensors*. A comma before a subscript represents covariant differentiation, which, in Cartesian coordinates is simply the partial derivative. Cartesian coordinates, however, are not suitable for integrating along streamlines. Therefore, the most promising solution methods involve the use of streamline coordinate systems, which can consolidate spatial dependence into a single variable. The solution can then be found by integrating along streamlines.

What has limited three-dimensional approaches until now is the fact that in 3D flows, vortex stretching can occur, so that $\boldsymbol{\omega}\cdot\nabla\boldsymbol{v} \neq 0$ in (3). Therefore, a substantial arsenal of methods available for two-dimensional flows is not available in three-dimensions (Majda & Bertozzi 2002). The added complexity of three-dimensional flow can be reduced, however, by taking a more general view of (3). In a general coordinate system, the two middle terms of (3) expand into the following expression:

$$\begin{aligned} & v^1\left(\frac{\partial\omega^i}{\partial x^1}+\Gamma^{\,i}_{1\,1}\omega^1+\Gamma^{\,i}_{2\,1}\omega^2+\Gamma^{\,i}_{3\,1}\omega^3\right)+v^2\left(\frac{\partial\omega^i}{\partial x^2}+\Gamma^{\,i}_{1\,2}\omega^1+\Gamma^{\,i}_{2\,2}\omega^2+\Gamma^{\,i}_{3\,2}\omega^3\right) \\ & +v^3\left(\frac{\partial\omega^i}{\partial x^3}+\Gamma^{\,i}_{1\,3}\omega^1+\Gamma^{\,i}_{2\,3}\omega^2+\Gamma^{\,i}_{3\,3}\omega^3\right)-\omega^1\left(\frac{\partial v^i}{\partial x^1}+\Gamma^{\,i}_{1\,1}v^1+\Gamma^{\,i}_{2\,1}v^2+\Gamma^{\,i}_{3\,1}v^3\right) \\ & -\omega^2\left(\frac{\partial v^i}{\partial x^2}+\Gamma^{\,i}_{1\,2}v^1+\Gamma^{\,i}_{2\,2}v^2+\Gamma^{\,i}_{3\,2}v^3\right)-\omega^3\left(\frac{\partial v^i}{\partial x^3}+\Gamma^{\,i}_{1\,3}v^1+\Gamma^{\,i}_{2\,3}v^2+\Gamma^{\,i}_{3\,3}v^3\right) \end{aligned} \tag{4}$$

where $\Gamma^{\,p}_{j\,k}$ are the connection coefficients, or Christoffel symbols of the second kind, given by:

$$\Gamma^{\,p}_{j\,k}=\frac{g^{pq}}{2}\left(\frac{\partial g_{jq}}{\partial x^k}+\frac{\partial g_{kq}}{\partial x^j}-\frac{\partial g_{jk}}{\partial x^q}\right).$$

Since $\Gamma^{\,p}_{j\,k}$ are not tensors, they do not vanish in some coordinate systems, such as axisymmetric coordinate systems, even though they vanish in a rectangular system. By the symmetry in the two lower indices of the Christoffel symbols however, 18 terms in (4) cancel out, so the expression in (4) reduces to

$$v^k\,\frac{\partial\omega^i}{\partial x^k}\;-\;\omega^k\,\frac{\partial v^i}{\partial x^k} \tag{5}$$

for any coordinate system with a symmetric connection. This is, of course, not limited to rectangular, cylindrical, and spherical coordinate systems but also includes non-orthogonal systems as well as arbitrary streamline coordinate systems. Replacing (4) with (5) can dramatically simplify computations in some numerical solution schemes, particularly when complicated coordinate systems are used. For Euler flows in particular, it may allow the use of more complex three-dimensional coordinate systems without additional computational overhead. Therefore, the statement made previously by the author (Morton 2004) that the vortex stretching term in (3) is zero if $v^2$ and $\omega^3$ are the only non-zero components of their respective tensors [setting $i=2$ and $j=3$ in (3)] and $v^2$ is independent of $x^3$, should be qualified as follows: *What remains of the vortex stretching term in* (3)*, after cancellation of terms containing Christoffel symbols*, is zero if $v^2$ and $\omega^3$ are the only non-zero components of their respective tensors and $v^2$ is independent of $x^3$ [set $i=2$ and $k=3$ in (5)].

The mathematical simplification above may also be useful in studies of electricity and magnetism, since the same nonlinear convective term arises there, but with the vorticity field replaced by the magnetic field strength (see, e.g., Hughes & Young 1989, chapter 6).

## 3. Vorticity Persistence in 3D Flows with Streamlines Fixed or in Steady Translation

As mentioned above, the simplification from (4) to (5) is valid in any coordinate system with a symmetric connection; therefore, it is also valid in streamline coordinates. Consider a streamline coordinate system in which the streamlines are fixed or translating with constant velocity. The velocity tensor in the streamline coordinate system will be denoted with overbars, and the nonzero component will be, say, $\overline{v}^2$. Then, when $i = 1$ or 3, the term on the right in (5) vanishes, and (3) becomes:

$$\frac{\partial \overline{\omega}^i}{\partial t} + \overline{v}^2 \frac{\partial \overline{\omega}^i}{\partial \overline{x}^2} = \nu(\overline{g}^{pk}\overline{\omega}^i_{,k})_{,p} \,. \qquad (i \neq 2) \qquad (6)$$

Since (3) can only be assumed valid in an inertial coordinate system, the streamlines comprising this coordinate system must be fixed or translating with constant velocity. The flow, however, may still be unsteady, as in the case of Oseen's vortex.

The two-dimensional solution of Oseen (1911) is given by:

$$\overline{\omega}^3 = \frac{\mathit{\Gamma}_O}{4\pi\nu\, t}\, e^{-(\overline{x}^1)^2/(4\nu\, t)} \,, \qquad \overline{\omega}^1,\, \overline{\omega}^2 = 0\,, \qquad (7)$$

$$\overline{v}^2 = \frac{\mathit{\Gamma}_O}{2\pi\,(\overline{x}^1)^2}\left(1 - e^{-(\overline{x}^1)^2/(4\nu\, t)}\right), \qquad \overline{v}^1,\, \overline{v}^3 = 0\,. \qquad (8)$$

Here, variables in the streamline coordinate system are denoted by overbars, $\mathit{\Gamma}_O$ is the total circulation, and $\overline{\omega}^k$ is the $k$th component of the vorticity tensor in the streamline coordinate system. In this case, the streamline coordinate system happens to be a cylindrical coordinate system; therefore, $\overline{x}^2$ is the spatial variable $\theta$ in the streamwise direction, and $\overline{v}^k = \mathrm{d}\overline{x}^k/\mathrm{d}t$ is the velocity tensor. The variable $\overline{x}^1$ in this case is the radius $R$ from the vortex center. Oseen's solution demonstrates how streamline coordinates simplify the mathematics.

In the case of Euler flow, (6) becomes:

$$\frac{\mathrm{d}\overline{\omega}^i}{\mathrm{d}t} = 0\,. \qquad (i \neq 2) \qquad (9)$$

Therefore, any component of the vorticity tensor not directed in the streamline direction must remain constant throughout the fluid particle motion. An equivalent result for unsteady, non-inertial streamlines is obtained in §6.

For steady Euler flow, (6) simplifies to:

$$\frac{\partial \overline{\omega}^i}{\partial \overline{x}^2} = 0\,. \qquad (i \neq 2) \qquad (10)$$

Since the directions $\overline{x}^1$ and $\overline{x}^3$ were arbitrary, integrating (10) shows that any component of the vorticity tensor not directed in the streamline direction must remain constant along the entire streamline.

These results are valid with or without swirl, and whether the flow is axisymmetric or not. The applicability of (9) is not limited to inviscid flow, since it is satisfied when $\nu\nabla^2\boldsymbol{\omega} = 0$, which includes both inviscid flows ($\nu = 0$) and flows with non-diffusive vorticity ($\nabla^2\boldsymbol{\omega} = 0$). It can happen that $\nabla^2\boldsymbol{\omega} = 0$ even when $\nabla^2\boldsymbol{v} \neq 0$, as in the case of the core in Hill's spherical vortex (Hill 1894). Interestingly, (10), which was derived as a special case of (9), can be valid in cases for which (9) is not. The unsteady Oseen solution, given by (7) and (8), is such a flow. Oseen's solution satisfies (10) because, as made clear by the use of the streamline coordinates $\overline{x}^k = (R, \theta, z)$ in (7), $\overline{\omega}^3$ is not a function of the streamwise direction, $\bar{x}^2$. However, (9) is not valid in the Oseen flow. In addition to (10), the Oseen solution separately satisfies the remaining terms of the unsteady vorticity equation, namely

$$\frac{\partial \boldsymbol{\omega}}{\partial t} \;=\; \nu\, \nabla^2 \boldsymbol{\omega}\,. \tag{11}$$

The expression in (5) is the Lie derivative, $\mathcal{L}_v$, of the vorticity tensor with respect to the fluid velocity tensor, $v$. It is interesting that many solutions that satisfy (3) also satisfy the following splitting of (3):

$$\mathcal{L}_v \boldsymbol{\omega} \;=\; 0\,, \qquad \frac{\partial \boldsymbol{\omega}}{\partial t} \;=\; \nu \nabla^2 \boldsymbol{\omega}\,. \tag{12}$$

As implied by the above discussion, it is too restrictive to view (12) as valid only for steady Euler flow because some flows obeying (12) are viscous, such as the core of Hill's spherical vortex. Likewise, referring to (12) as valid for "flows with non-diffusive vorticity," wherein $\nu\nabla^2\boldsymbol{\omega} = 0$ either because $\nu = 0$ or $\nabla^2\boldsymbol{\omega} = 0$, is also too restrictive because, as in the case of Oseen's unsteady vortex, the vorticity may be diffusive and yet (12) remain satisfied. Therefore, an appropriate label for flows satisfying (12) might be: "flows with streamlines that are fixed in some steadily translating coordinate system."

## 4. A Close-up View of Vortex Stretching

To see the two distinct contributions to the vortex stretching term, consider the Euler flows depicted in Figure 1, in which an unsteady (stretching) vortex is shown in (a) and a steady axisymmetric vortex in (b). For both cases, the vortex stretching term of (3) is

$$\bar{\varpi}^3 \bar{v}^i{}_{,3} = \bar{\varpi}^3 \left( \frac{\partial \bar{v}^i}{\partial \bar{x}^3} + \Gamma^{\,i}_{p3} \bar{v}^p \right). \tag{13}$$

In both cases, $\bar{v}^2$ is the only non-zero velocity component. Therefore, for a flow such as that in Figure 1a, if we write (13) for the streamline direction by setting $i = 2$, the first term in parentheses is non-zero since $\bar{v}^2$ changes with $\bar{x}^3$. For the flow in Figure 1b, on the other hand, both terms on the right side of (13) are zero when $i = 2$. However, when $i = 3$, the vortex stretching term is not zero even though $\bar{v}^3 = 0$. This is because the second term in parentheses is non-zero. For example, for Hill's spherical vortex, with $i = 3$,

$$\tilde{\Gamma}^{\,3}_{23} = \frac{\tilde{g}^{33}}{2}\frac{\partial \tilde{g}_{33}}{\partial \tilde{x}^3},$$

where $\tilde{x}^3$ is the azimuthal coordinate in a spherical coordinate system, and $\tilde{g}^{33} = 1/\tilde{g}_{33} = 1/(r\sin\theta)^2$ (see §7 for more discussion of this point). As shown in §2, this non-zero term ultimately does not enter into the vorticity equation because it is exactly cancelled by an equal and opposite contribution elsewhere.

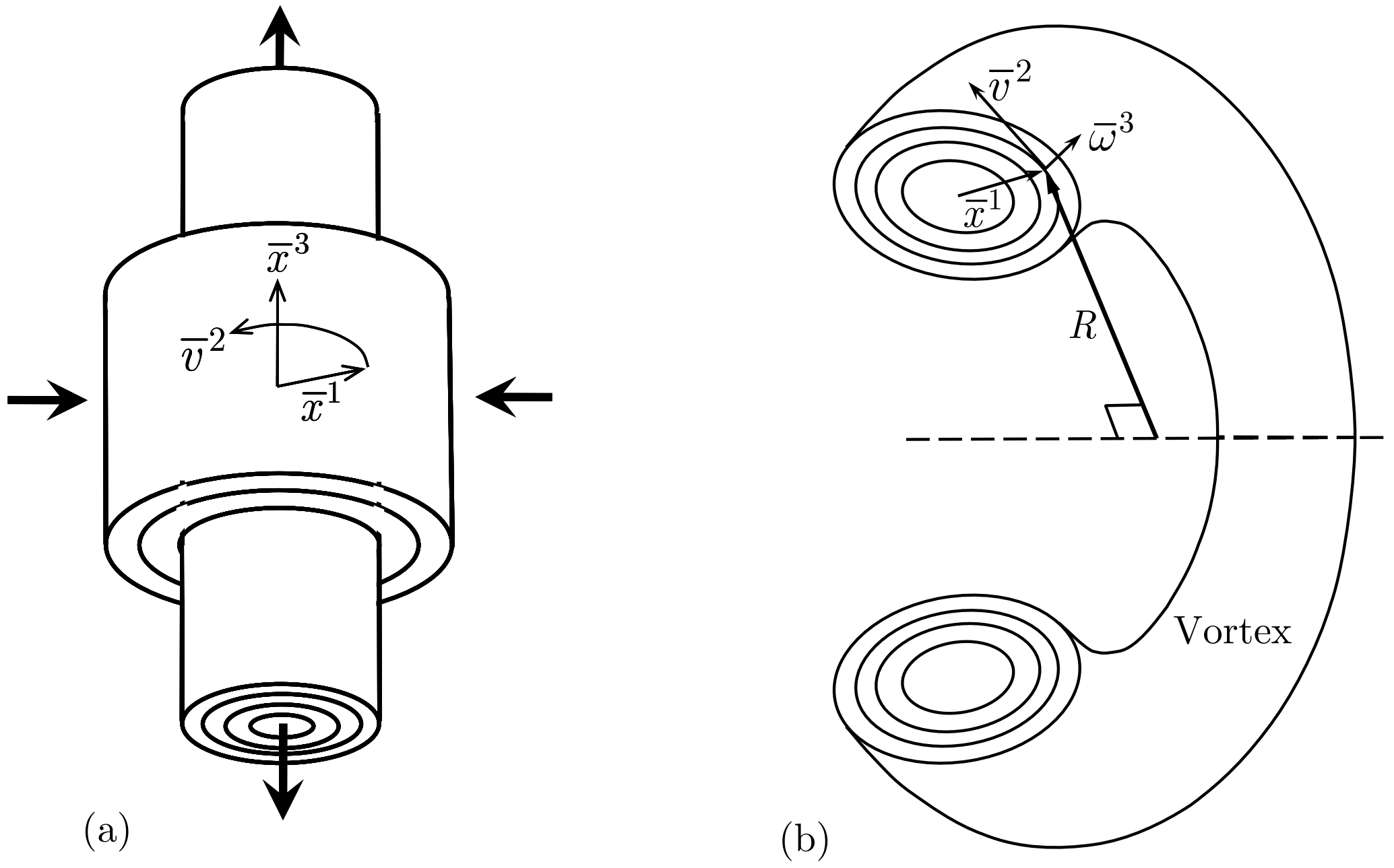

Figure 1. Schematic showing the two causes of a non-vanishing vortex stretching term. In (a), the term $\bar{\varpi}^3\, \partial \bar{v}^2/\partial \bar{x}^3$ in (13) is non-zero; in (b), the term $\bar{\varpi}^3 \Gamma^3_{22} \bar{v}^2$ is non-zero.

## 5. Continuity Equation

The unsteady continuity equation, written for a general coordinate system, can be derived from the Reynolds transport theorem, as follows:

$$\underbrace{\frac{\mathrm{d}}{\mathrm{d}t}\int \rho\sqrt{g}\,\mathrm{d}V\!\!\!\!/}_{\text{fluid particle frame}} \;=\; \underbrace{\int\left(\frac{\partial}{\partial t}\;\rho\sqrt{g} \;+\; \rho\sqrt{g}\,v^k{}_{,k}\right)\mathrm{d}V\!\!\!\!/}_{\text{alternate frame}} \;=\; 0\,.$$

Here, $\sqrt{g}$ is the Jacobian determinant of the coordinate system. Since the integrands must be zero, the middle term above gives the general continuity equation (compressible in this case) in an arbitrary coordinate frame:

$$\frac{\partial}{\partial t}(\rho\sqrt{g}) \;+\; \mathcal{L}_v(\rho\sqrt{g}) \;=\; 0\,. \tag{14}$$

For the case of a streamline coordinate system "frozen" in a steadily translating frame, the unsteady term above simplifies to $\partial\rho/\partial t$.

## 6. Vorticity Persistence in 3D Flows with General, Time-dependent Streamlines

As previously noted, (3) is only applicable to an inertial coordinate system, and this poses a problem if the coordinate system is to coincide with general unsteady streamlines. In such a coordinate system, the metric tensor will be time-dependent; therefore, the Christoffel symbols will be also. At every instant of time, however, the cancellation in §2 still occurs, since the Christoffel symbols only involve spatial relationships.

Consider (3) without the viscous term:

$$\frac{\partial \omega^i}{\partial t} \;+\; v^k\,\frac{\partial \omega^i}{\partial x^k} \;-\; \omega^k\,\frac{\partial v^i}{\partial x^k} \;=\; 0\,. \tag{15}$$

To avoid the difficulty imposed by the restriction to inertial frames, the vorticity equation (3) will be written in terms of a non-inertial coordinate system. In a manner analogous to the coordinate-independent definition of differentiation afforded by the covariant derivative, a general definition of temporal differentiation will be constructed so that (15) may be written in terms of non-inertial coordinates. Temporal differentiation of the contravariant first-order tensor $\omega^i$ can be expressed in terms of components $\bar{\omega}^k$ of the vorticity tensor referenced to a *non-inertial* coordinate system, as follows:

$$\frac{\mathrm{d}\omega^i}{\mathrm{d}t} \;=\; \frac{\mathrm{d}}{\mathrm{d}t}\left(\frac{\partial x^i}{\partial \bar{x}^k}\,\bar{\omega}^k\right). \tag{16}$$

Performing the indicated differentiation with respect to time gives:

$$\frac{\mathrm{d}\omega^i}{\mathrm{d}t} = \frac{\mathrm{d}}{\mathrm{d}t}\left(\frac{\partial x^i}{\partial \overline{x}^k}\right)\overline{\omega}^k + \frac{\partial x^i}{\partial \overline{x}^k}\frac{\mathrm{d}\overline{\omega}^k}{\mathrm{d}t}.$$

Replacing $\overline{\omega}^k$ in the middle term with $\omega^j\ \partial\overline{x}^k/\partial x^j$ gives:

$$\frac{\mathrm{d}\omega^i}{\mathrm{d}t} - \frac{\mathrm{d}}{\mathrm{d}t}\left(\frac{\partial x^i}{\partial \overline{x}^k}\right)\frac{\partial \overline{x}^k}{\partial x^j}\omega^j = \frac{\partial x^i}{\partial \overline{x}^k}\frac{\mathrm{d}\overline{\omega}^k}{\mathrm{d}t}. \tag{17}$$

A temporal differentiation is now defined as the left side of (17):

$$\omega^i_{\ ,T} \equiv \frac{\mathrm{d}\omega^i}{\mathrm{d}t} - \frac{\mathrm{d}}{\mathrm{d}t}\left(\frac{\partial x^i}{\partial \overline{x}^k}\right)\frac{\partial \overline{x}^k}{\partial x^j}\omega^j, \tag{18}$$

where the subscript "$T$" denotes "time" and is capitalized to prevent confusion with an index. Now (17) may be abbreviated as:

$$\omega^i_{\ ,T} = \frac{\partial x^i}{\partial \overline{x}^k}\frac{\mathrm{d}\overline{\omega}^k}{\mathrm{d}t}. \tag{19}$$

The middle term of (17) treats the temporal characteristics in the same manner that the Christoffel symbols treat spatial characteristics.

Reversing the order of differentiation in the definition (18) gives:

$$\omega^i_{\ ,T} = \frac{\mathrm{d}\omega^i}{\mathrm{d}t} - \frac{\partial}{\partial \overline{x}^k}\left(\frac{\mathrm{d}x^i}{\mathrm{d}t}\right)\frac{\partial \overline{x}^k}{\partial x^j}\omega^j. \tag{20}$$

The derivative in the parentheses above is $v^i$; therefore, the definition, as expressed in (20), simplifies to

$$\omega^i_{\ ,T} = \frac{\mathrm{d}\omega^i}{\mathrm{d}t} - \frac{\partial v^i}{\partial x^j}\omega^j. \tag{21}$$

Therefore, the vorticity equation in (15) may be written simply as

$$\omega^i_{\ ,T} = 0. \tag{22}$$

Writing now the definition (21) for the non-inertial coordinate system, we have:

$$\overline{\omega}^k_{\ ,T} = \frac{\mathrm{d}\overline{\omega}^k}{\mathrm{d}t} - \frac{\partial \overline{v}^k}{\partial \overline{x}^j}\overline{\omega}^j, \tag{23}$$

with no claim that it represents the vorticity equation in a non-inertial coordinate system. Suppose that this non-inertial coordinate system coincides with an unsteady streamline coordinate system. If $\overline{x}^2$ is the streamline direction, we may set $\overline{v}^k = 0$ when $k \neq 2$, so that (23), written for directions other than the streamline direction, becomes:

$$\overline{\omega}^k_{\ ,T} = \frac{\mathrm{d}\overline{\omega}^k}{\mathrm{d}t}, \qquad (k \neq 2) \tag{24}$$

Substituting (24) into (19) yields:

$$\omega^i_{\ ,T} = \frac{\partial x^i}{\partial \overline{x}^k}\overline{\omega}^k_{\ ,T}. \qquad (k \neq 2) \tag{25}$$

This is the transformation required of a contravariant first-order tensor, but it is only valid on the two-dimensional manifold traversed by $\overline{x}^1$ and $\overline{x}^3$. The right side of (25) is a summation of 2 terms, corresponding to $k = 1$ and 3. According to (22), which is the vorticity equation in the inertial system, this summation must be zero. However, the factors $\partial x^i / \partial \bar{x}^k$ in (25) are, in general, independent and non-zero; therefore, $\bar{\omega}^1{}_{,T}$ and $\bar{\omega}^3{}_{,T}$ must both be zero; that is

$$\bar{\omega}^k{}_{,T} \; = \; 0 \, . \qquad\qquad (k \neq 2) \qquad\qquad (26)$$

The proviso above that $k \neq 2$ simply excludes the streamline direction from the relation. According to (24), (26) may also be written:

$$\frac{\mathrm{d}\bar{\omega}^k}{\mathrm{d}t} \; = \; 0 \, . \qquad\qquad (k \neq 2) \qquad\qquad (27)$$

Therefore, as in the 2-D Euler case, in any 3-D flow satisfying (15), all components of the vorticity tensor in directions other than the streamline direction are conserved along streamlines. Since the directions $\bar{x}^1$ and $\bar{x}^3$ were arbitrary, this result is valid in any coordinate system, both inertial and non-inertial, provided its transformation to inertial coordinates is non-singular.

If the temporal derivative defined by (18), or equivalently by (21), is written as:

$$(\;)_{,T} \; \equiv \; \frac{\partial}{\partial t} \; + \; \mathcal{L}_v \, ,$$

then the continuity equation (14) and the vorticity equation given in (15) may both be written simply as $(\;)_{,T} \; = \; 0$, with $\rho\sqrt{g}$ and $\boldsymbol{\omega}$ as their respective arguments.

## 7. A Unifying Statement about Vorticity Persistence

In axisymmetric flows without swirl, a spherical coordinate system is almost as convenient as a streamline coordinate system because the azimuthal variables (such as vorticity) coincide for the two systems. Let $\tilde{x}^k = (r, \theta, \phi)$ represent components of a spherical coordinate system, $\tilde{\omega}(3)$ the azimuthal component of the *physical* vorticity in this spherical system, and $R = r \sin\theta$ the distance to the symmetry axis (see Figure 1b). The quantity $\tilde{\omega}(3)$ is actually the azimuthal component of the physical vorticity for any axisymmetric flow. The quantity $\tilde{\omega}(3)/R$ is one that has been employed in several past studies. For vortex rings of small cross-section, Fraenkel (1970) proved that if the ratio $\tilde{\omega}(3)/R$ is constant along streamlines, then steady solutions exist. Fraenkel's method of proof relied upon the nearly "two-dimensional nature of the flow in the neighborhood of a small cross-section." Norbury (1972) remarked that it had long been thought that certain vortex rings of small cross-section are steady if the above ratio is constant throughout their cores. Norbury (1973) assumed this ratio to be constant in obtaining his well-known family of solutions. The ratio $\tilde{\omega}(3)/R$ has been referred to as a "vorticity constant" (Norbury 1973) as well as a "vorticity density" (Mohseni 2001; Mohseni & Gharib 1998). Hunt and Eames (2002) noted that during axisymmetric vortex stretching in a straining flow, this ratio is conserved.

The ratio referred to above is actually the azimuthal component of the vorticity *tensor* of the flows referenced in the above studies, since the physical component of vorticity is $\tilde{\omega}(3) = \sqrt{\tilde{g}_{33}}\tilde{\omega}^3 = (r \sin\theta)\,\tilde{\omega}^3 = R\tilde{\omega}^3$ (Morton 2004). Here, $\tilde{\omega}^3$ is the azimuthal component of the vorticity *tensor* in a spherical coordinate system (and, for the flows mentioned in the above paragraph, a streamline coordinate system as well). The tensor component $\tilde{\omega}^3$ is constant along streamlines. However, the result in (10), and more generally (27), indicates that *all* components of the vorticity tensor normal to the streamline direction (not only the azimuthal component in axisymmetric flow) are conserved in any flow satisfying $\nu\nabla^2\boldsymbol{\omega} = 0$ (not only in Euler rings of small cross-section and not only in planar or axisymmetric flows).

## 8. Conclusion

The vorticity equation for arbitrary coordinate systems is simplified by a complete cancellation of all Christoffel symbols from the nonlinear convective terms. Whereas vorticity has been known for more than a century to persist in 2-D Euler flow, in the last few decades researchers have established that for inviscid axisymmetric Euler rings, it is the quantity $\omega/R$, rather than $\omega$, that persists. It was shown herein that this apparent contrast vanishes when one states that: in a streamline coordinate system, components of the vorticity *tensor* in directions other than the streamline direction are conserved along streamlines if the condition $\nu\nabla^2\boldsymbol{\omega} = 0$ is satisfied. Therefore, the known persistence of vorticity in two-dimensional Euler flow has been extended to a class of three-dimensional Euler flows in which a streamline coordinate system exists and the streamline velocity component is independent of one transverse coordinate. Another consequence of the simplification of the vorticity equation is the finding that what remains of the non-linear convective terms, after the mathematical simplification, constitutes the Lie derivative of the vorticity tensor with respect to fluid velocity.